\documentclass[aps,preprint,amsmath,amssymb]{revtex4}
\usepackage{graphicx}
\newcommand{\nn}{\nonumber}
\newcommand{\bd}{\begin{document}}
\newcommand{\ed}{\end{document}}
\newcommand{\bc}{\begin{center}}
\newcommand{\ec}{\end{center}}
\newcommand{\be}{\begin{eqnarray}}
\newcommand{\ee}{\end{eqnarray}}
\newcommand{\ba}{\begin{array}}
\newcommand{\ea}{\ed{array}}
\newcommand{\strich}[1]{#1  \! \! \slash}
\newcommand{\eqn}{\global\def\theequation}
\newcommand{\sw}{sin^2 \theta_W}
\newcommand{\fbd}{f_B}
\renewcommand{\thefootnote}{\alph{footnote}}
\newcommand{\se}{\section}
\newcommand{\sse}{\subsection}
\newcommand{\bi}{\bibitem}
\def\figcap{\section*{Figure Captions\markboth
     {FIGURECAPTIONS}{FIGURECAPTIONS}}\list
     {Figure \arabic{enumi}:\hfill}{\settowidth\labelwidth{Figure 999:}
     \leftmargin\labelwidth
     \advance\leftmargin\labelsep\usecounter{enumi}}}
\let\endfigcap\endlist \relax
\def\reflist{\section*{References\markboth
     {REFLIST}{REFLIST}}\list
     {[\arabic{enumi}]\hfill}{\settowidth\labelwidth{[999]}
     \leftmargin\labelwidth
     \advance\leftmargin\labelsep\usecounter{enumi}}}
\let\endreflist\endlist \relax

\begin{document}
\title
{\Large
 \bf
Analysis  of
 $K^+ \to e^+ \nu_{e}\gamma$ in light-front quark model and
chiral perturbation theory of order
$p^6$ }

\author{ \bf \large Chuan-Hung Chen$^{1,2}$, Chao-Qiang Geng$^{3}$
  and Chong-Chung Lih$^{4}$\\
 }

\affiliation{  $^{1}$Department of Physics, National Cheng-Kung
University, Tainan 701, Taiwan \\
$^{2}$National Center for Theoretical Sciences, Hsinchu 300,Taiwan\\
$^{3}$Department of Physics, National Tsing-Hua University,
Hsinchu
300, Taiwan  \\
$^4$General Education Center, Tzu-Chi  College of Technology,
Hualien 970, Taiwan
 }

\date{\today}

\begin{abstract}
Within the  frameworks of the light-front quark model (LFQM) and chiral perturbation theory (ChPT)
of  $O(p^6)$,
we reevaluate the  form factors of the $K^+ \to \gamma$ transition.
We  use these form factors to study the decay
of $K^+ \to e^+ \nu_{e}\gamma$, which is dominated by the structure dependent  contribution.
We show the differential 
decay branching ratio as a function of $x=2E_\gamma/m_K$, where $E_\gamma$ ($m_K$)
is the photon energy (kaon mass).
Explicitly, we find that, in the standard model
with the cut of $x=0.01$ ($0.1$),
 the decay branching ratio of $K^+ \to e^+ \nu_{e}\gamma$ is
 $1.54\ (1.44)\times 10^{-5}$ and $1.57\ (1.47)\times 10^{-5}$
 in the LFQM and ChPT, respectively.

\end{abstract}

\maketitle %

\se{Introduction}

Experimentally, both
decays of $K^+\to e^+\nu_e$ and $\mu^+\nu_\mu$ have been precisely measured with the decay branching ratios
being $(1.55\pm0.05)\times 10^{-5}$ and $(63.44\pm0.14)\times 10^{-2}$ \cite{pdg}, respectively.
The smallness of the electron mode can be easily understood as it is helicity suppressed with the suppression factor
of $m^2_e/m_\mu^2\sim 2\times 10^{-5}$ in comparison with the muon mode.
For the corresponding radiative decays of $K^+\to \ell^+\nu_{\ell}\gamma\ (\ell=e,\mu)$,
it is known that they receive two types of contributions:  ``inner bremsstrahlung'' (IB) and ``structure-dependent'' (SD)
\cite{GW,Bryman}. For the decay of $K^+\to e^+\nu_e\gamma$, while the IB contribution is still helicity suppressed and
contains the electromagnetic coupling constant $\alpha$ as well,
 the SD part gives the dominant contribution to the decay rate as
 it is free of the helicity suppression. Similarly, the SD contribution is also important to the decay of
 $K^+\to \mu^+\nu_{\mu}\gamma$ \cite{Bijnens93}.

  In the standard model (SM), the decay amplitude
 of the SD part  involves
 vector and axial-vector hadronic currents, which can be parametrized in terms of the vector form factor $F_V$ and
 axial-vector form factor $F_A$, respectively.
 However, the experimental determinations on these form factors are
  poorly given and model-dependent \cite{75,79,E787}. In particular, the experimental results on the 
  decay rate of $K^+\to e^+\nu_e\gamma$ in Ref. \cite{75,79,E787} were based on the assumption of $F_V$ and $F_A$ being some constant values in
   the chiral perturbation theory (ChPT)
at $O(p^4)$ \cite{Bijnens93}.
   In the ongoing data analysis of the E949 experiment at BNL,
 more precision measurements on the decay of $K^{+}\to e^{+}\nu_{e}\gamma$ are expected \cite{E949}
 and thus, the model-independent extractions of the SD form factors are possible.
 Theoretical calculations of $F_V$ and $F_A$ in the $K^+\to \gamma$ transition
have been previously done
 in the 
ChPT
at $O(p^4)$ \cite{Bijnens93}
and $O(p^6)$ \cite{mod3,mod4}.
However, the results of the ChPT at $O(p^6)$ \cite{mod4} have not been fully applied to
the decay of $K^{+}\to e^{+}\nu_{e}\gamma$ yet. Moreover, it is important if we could obtain information on $F_{V,A}$ in some QCD model other than the ChPT. For this purpose, in the present  study we will also evaluate
  $F_{V,A}$ in the light front quark model (LFQM).
We will use the form factors in both  ChPT and LFQM to examine the decay of $K^{+}\to e^{+}\nu_{e}\gamma$.

This paper is organized as follows:  We present
the relevant formulas for the matrix elements and form factors  for the decay of $K^{+}\to e^{+}\nu_{e}\gamma$ in Sec.~II.
In particular, we study
 the transition form factors of $K^+\to\gamma$ in the ChPT of $O(p^6)$ and LFQM.
In Sec.~III, we describe  the differential decay rate of $K^{+}\to e^{+}\nu_{e}\gamma$.
In Sec.~IV, we show
our numerical results on the form factors and the
decay  branching ratio
in both ChPT and LFQM.
We will also illustrate  the differential decay branching ratio as a function of  $x=2E_\gamma/m_K$, where $E_\gamma$ and
$m_K$ are the photon energy and kaon mass, respectively.
We give our conclusions in Sec.~V.

\se{Matrix elements and form factors}

In the SM,
 the amplitude of the decay $K^{+}\to e^{+}\nu_{e}\gamma$
 ($K^+_{e2\gamma}$) can be written
in terms of
IB and SD contributions, given by
 \cite{Bryman,Bijnens93,ffdef,Chen1}
\be
M&=&M_{IB}+M_{SD}, \nonumber
\\
M_{IB}&=&i e{G_F \over \sqrt{2}} sin\theta_c F_K m_{e}
 \epsilon^*_{\alpha}K^{\alpha},
\nn
\\
M_{SD}&=&-i e{G_F \over \sqrt{2}} sin\theta_c \epsilon^*_{\mu}L_{\nu}H^{\mu\nu},
\label{eqn:sd}
\ee
where
\be
K^{\alpha}&=&\bar{u}(p_{\nu})(1+\gamma_5)\left ({p_K^{\alpha} \over p_K\cdot q}
-{2 p^{\alpha}_{e}+ \not\!q \gamma^{\alpha} \over 2 p_{e}\cdot q}
\right)v(p_{e}),
\nonumber \\
L_{\nu}&=&\bar{u}(p_{\nu})\gamma_{\nu}(1-\gamma_5)v(p_{e}), \nonumber\\
H^{\mu\nu}&=&{F_A \over m_K}\left(-g^{\mu\nu}p_K\cdot q+p_K^{\mu}q^{\nu}\right)+
i{F_V \over m_K}\epsilon^{\mu\nu\alpha\beta}q_{\alpha}p_{K\beta}\,,
\label{27}
\ee
$\epsilon_{\alpha}$ is the photon polarization vector,
$p_K$, $p_{\nu}$, $p_{e}$, and $q$ are the four-momenta of
$K^+$, $\nu_e$, $e^+$, and $\gamma$,
and $F_K$ and $F_{A(V)}$
are the $K$ meson decay constant
and  the axial-vector
(vector) form factor corresponding to the axial-vector (vector) part of the weak currents,
 defined by
\be
\langle\, 0|\bar{s}\gamma^{\mu}\gamma_5 u|K^+(p_K) \,\rangle &=&-iF_Kp_K^{\mu},
\nn\\
\langle\gamma(q) |\bar{u}\gamma^{\mu }\gamma _{5}s|K(p_K) \,\rangle &=&
e{\frac{F_{A}}{%
m_{K}}}\left[ (p\cdot q) \epsilon ^{* \mu}
-(\epsilon ^{*}\cdot p)q^{\mu }\right] ,
\nonumber\\
\langle\gamma(q) |\bar{u}\gamma^{\mu }s|K(p_K) \,\rangle &=&
ie{\frac{F_{V}}{m_{K}}}\varepsilon^{\mu \alpha \beta \nu }
\epsilon_{*\alpha }q_{\beta }p_\nu \, ,
\label{4}
\ee
respectively,
with $p=p_K -q$ being the transfer momentum.
We note that ${\cal M}_{IB}$  in Eq. (\ref{eqn:sd}) is suppressed due to the small
electron mass $m_{e}$.
In the decay of $K^{+}\to e^{+}\nu_{e}\gamma$,
the form factors $F_{A,V}$ in Eq.  (\ref{4}) are the analytic functions of $p^{2}=(p_K -q)^2$ in
 the physical allowed region, given by
 \be
m_{e}^{2}\leq p^{2}\leq m_{K}^{2}\,.
\ee
In the following discussion, we will first 
summarize the formulas for $F_{V,A}$ in the ChPT and then  study these
form factors
in the LFQM. We note that similar calculations for the $P\to \gamma\ (P=K^0,D,B)$ transitions in the LFQM have
been performed in Refs. \cite{modLF,lf1,mod1}.

\sse{Chiral Perturbation Theory}

The chiral Lagrangians contain both normal and anomalous parts.
At orders 
of $p^m$,
the non-anomalous and anomalous Lagrangians of ${\cal L}_n^{(m)}$ and  ${\cal L}_a^{(m)}$ relevant to the
$K^+_{e 2\gamma}$ decay  are given by  \cite{mod4}
\be
{\cal L}_{n}^{(2)} &=&\frac{F^{2}}{4}Tr(D_{\mu }UD^{\mu }U^{\dagger })+\frac{F^{2}}{%
4}Tr(\chi U^{\dagger }+U\chi ^{\dagger })\,,
\nn
\ee
  \be
  {\cal
L}_{n}^{(4)} &=&L _{1}\left[ Tr(D_{\mu }UD^{\mu }U^{\dagger
})\right] ^{2}+L _{2}Tr(D_{\mu }UD_{\nu }U^{\dagger })Tr(D^{\mu
}UD^{\nu }U^{\dagger })  \nonumber \\ &&+L _{3}Tr(D_{\mu }UD^{\mu
}U^{\dagger }D_{\nu }UD^{\nu }U^{\dagger }) +L
_{4}Tr(D_{\mu }UD^{\mu }U^{\dagger })Tr(\chi U^{\dagger }+U\chi
^{\dagger })  \nonumber \\ &&+L _{5}Tr(D_{\mu }UD^{\mu }U^{\dagger
}(\chi U^{\dagger }+U\chi
^{\dagger }))+L _{6}\left[ Tr(\chi U^{\dagger }+U\chi ^{\dagger })%
\right] ^{2}  \nonumber \\ &&+L _{7}\left[ Tr(\chi ^{\dagger
}U-U^{\dagger }\chi )\right] ^{2}+L _{8}Tr(\chi U^{\dagger }\chi
U^{\dagger }+U\chi ^{\dagger }U\chi ^{\dagger })  \nonumber \\
&&+iL _{9}Tr(L_{\mu \nu }D^{\mu }UD^{\nu }U^{\dagger }+R_{\mu \nu
}D^{\mu }U^{\dagger }D^{\nu }U)+L _{10}Tr(L_{\mu \nu }UR_{\mu \nu
}U^{\dagger })\,,  \label{P4}
\nn
\ee
\be
{\cal L}^{(6)}_{n}&=&y_{17}\langle \chi_{+}h_{\mu\nu}h^{\mu\nu} \rangle+
y_{18}\langle \chi_{+} \rangle \langle h_{\mu\nu}h^{\mu\nu} \rangle+
y_{81}\langle \chi_{+}f_{+\mu\nu}f_{+}^{\mu\nu} \rangle \nonumber\\
&&+y_{82}\langle \chi_{+} \rangle \langle f_{+\mu\nu}f_{+}^{\mu\nu} \rangle+
iy_{83}\langle f_{+\mu\nu} \left\{ \chi_{+},u^{\mu}u^{\nu} \right\} \rangle+
iy_{84}\langle \chi_{+} \rangle \langle f_{+\mu\nu} u^{\mu}u^{\nu} \rangle \nonumber\\
&&+iy_{85}\langle f_{+\mu\nu} u^{\mu} \chi_{+} u^{\nu} \rangle+
iy_{100}\langle f_{-\mu\nu} [ f_{-}^{\nu\rho},h^{\mu}_{\rho}] \rangle+
y_{102}\langle \chi_{+} f_{-\mu\nu} f_{-}^{\mu\nu} \rangle \nonumber\\
&&+y_{103}\langle \chi_{+} \rangle \langle f_{-\mu\nu} f_{-}^{\mu\nu} \rangle+
y_{104}\langle f_{+\mu\nu} [ f_{-}^{\mu\nu},\chi_{-}] \rangle+
y_{109}\langle \bigtriangledown_{\rho} f_{-\mu\nu}
\bigtriangledown^{\rho}f_{-}^{\mu\nu} \rangle \nonumber\\
&&+y_{110}\langle \bigtriangledown_{\rho} f_{+\mu\nu}
[ h^{\nu\rho},u^{\nu}] \rangle+....\,,
\label{p6}
\ee
and \cite{ch1,ch2}
\be
{\cal L}_{a}^{(4)}&=&-\frac{1}{16\pi^2}\epsilon^{\mu\nu\alpha\beta}
Tr\left(U \partial_{\mu} U^{+}\partial_{\nu}U\partial_{\alpha}U^{+}l_{\beta}-
U^{+} \partial_{\mu} U\partial_{\nu}U^{+} \partial_{\alpha}U r_{\beta} \right)
\nonumber\\
&&-\frac{i}{16\pi^2}\epsilon^{\mu\nu\alpha\beta}
Tr\left(\partial_{\mu} U^{+} \partial_{\nu} l_{\alpha} U r_{\beta}-
\partial_{\mu} U \partial_{\nu} r_{\alpha} U^{+} l_{\beta} \right)  \nonumber\\
&&+U \partial_{\mu} U^{+}\left( l_{\nu} \partial_{\alpha} l_{\beta}
+\partial_{\nu} l_{\alpha} l_{\beta} \right) \,,
\nn
\label{anomp4}
\\
{\cal L}_{a}^{(6)}&=&i C_{7} \epsilon^{\mu\nu\alpha\beta}
\langle \chi_{-} f_{+\mu\nu} f_{+\alpha\beta} \rangle +
i C_{11} \epsilon^{\mu\nu\alpha\beta}
\langle \chi_{-} [f_{+\mu\nu}, f_{-\alpha\beta}] \rangle  \nonumber\\
&+& C_{22} \epsilon^{\mu\nu\alpha\beta}
\langle u_{-} \left\{\bigtriangledown_{\gamma} f_{+\gamma\nu},
f_{+\alpha\beta} \right\} \rangle+....\,,
\label{anomp6}
\ee
respectively,
where $F$ is the meson decay constant in the chiral limit,
$L _{i}$,  $y_{j}$ and $C_k$ are unrenormalized coupling constants,
$U$ is the unitary matrix, parametreized by
\be
U = \exp\left[i\frac{\sqrt{2}}{F}\left(
\begin{array}{ccc}
\frac{\pi^{0}}{\sqrt{2}}+\frac{\eta}{\sqrt{6}} & \pi^{+} & K^{+} \\
\pi^{-} & -\frac{\pi^{0}}{\sqrt{2}}+\frac{\eta}{\sqrt{6}} & K^{0} \\
K^{-} & K^{0} & -\frac{2\,\eta}{\sqrt{6}}
\end{array}
\right)\right]\,,
\label{U}
\ee
$L_{\mu \nu }$ and $R_{\mu \nu }$ are the field-strength tensors of
external sources, given by
\be
L_{\mu \nu }&=&\partial _{\mu }\ell _{\nu }-\partial _{\nu }\ell _{\mu }-i
\left[ \ell _{\mu },\ell _{\nu }\right] \,,  \nonumber \\
R_{\mu\nu }&=&\partial _{\mu }r_{\nu }-\partial _{\nu }r_{\mu }-i\left[
r_{\mu },r_{\nu }\right]\,,
\ee
and the definitions of all other fields can be found in Ref. \cite{mod4}.

 From the chiral Lagrangians in Eqs.~(\ref{p6}) and (\ref{anomp6}),
one obtains the tree and loop contributions to $F_{V}$ at $O(p^6)$ for the $K^+_{e 2\gamma}$ decay to be \cite{mod3,mod4}
\be
F_{V}(p^2)&=&\frac{m_{K}}{4\sqrt{2}\pi^{2}F_K}\bigg\{1-\frac{256}{3}\pi^{2}m_{K}^{2}C_{7}^{r}
+256\pi^{2}(m_{K}^{2}-m_{\pi}^{2})C_{11}^{r}+\frac{64}{3}\pi^{2} p^{2} C_{22}^{r}
\nonumber\\
&-&\frac{1}{16 \pi^{2}(\sqrt{2}F_{K})^{2}}\bigg[\frac{3}{2}m_{\eta}^{2}
\ln \left(\frac{m_{\eta}^{2}}{\mu^{2}}\right)+\frac{7}{2} m_{\pi}^{2}
\ln \left(\frac{m_{\pi}^{2}}{\mu^{2}}\right)
+3m_{K}^{2} \ln \left(\frac{m_{K}^{2}}{\mu^{2}}\right)
\nonumber\\
&-&2\int \left[x m_{\pi}^{2}+(1-x) m_{K}^{2}-x(1-x)p^{2}\right]
\ln\left(\frac{x m_{\pi}^{2}+(1-x) m_{K}^{2}-x(1-x)p^{2}}{\mu^{2}}\right)dx
\nonumber\\
&-&2\int \left[x m_{\eta}^{2}+(1-x) m_{K}^{2}-x(1-x)p^{2}\right]
\ln\left(\frac{x m_{\eta}^{2}+(1-x) m_{K}^{2}-x(1-x)p^{2}}{\mu^{2}}\right)dx
\nonumber\\
&-&4\int m_{\pi}^{2}\,
\ln\left(\frac{m_{\pi}^{2}}{\mu^{2}}\right)dx
\bigg]\bigg\}\,,
\label{fvk}
\ee
where  the wave function and decay constant renormalizations have been included and
$C^{r}_{i}$ are
the renormalized coefficients.
 From 
  Eq.~(\ref{p6}), the tree and loop contributions to  $F_A$ of $O(p^6)$
 lead to \cite{mod4}
\be
F_{A}(p^2)&=&\frac{4 \sqrt{2}m_{K}}{F_{K}}(L _{9}^{r}+L _{10}^{r})+
\frac{m_{K}}{6F_{K}^{3}(2\pi)^{8}}[142.65 (m_{K}^{2}-p^{2})-198.3]
\nonumber \\
&-&\frac{m_{K}}{4\sqrt{2}F_{K}^{3}\pi^{2}}\left\{ (4L
_{3}^{r}+7L _{9}^{r}+7L _{10}^{r})m_{\pi }^{2}\ln \left(\frac{m_{\pi }^{2}}{m_{\rho }^{2}%
}\right)+3\,(L _{9}^{r}+L _{10}^{r})m_{\eta }^{2}\ln \left(\frac{m_{\eta }^{2}}{%
m_{\rho }^{2}}\right)\right.  \nonumber \\ &&\left. +2\,(8L
_{1}^{r}-4L _{2}^{r}+4L _{3}^{r}+7L _{9}^{r}
+7L _{10}^{r})m_{K}^{2}\ln \left(\frac{m_{K}^{2}}{m_{\rho }^{2}}%
\right)\right\} \nonumber \\
&-&\frac{4\sqrt{2}m_{K}}{3F_{K}^{3}}\left\{ 2m_{\pi
}^{2}(18y_{18}^{r}-2y_{81}^{r}-6y_{82}^{r}
+2y_{83}^{r}+3y_{84}^{r}-y_{85}^{r}+6y_{103}^{r})\right. \nonumber
\\ &&+2m_{K}^{2}(18y_{17}^{r}+36y_{18}^{r}-4y_{81}^{r}
-12y_{82}^{r}+4y_{83}^{r}+6y_{84}^{r}+4y_{85}^{r}-3y_{100}^{r}
\nonumber \\ &&\left.
+6y_{102}^{r}+12y_{103}^{r}-6y_{104}^{r}+3y_{109}^{r})
+\frac{3}{2}(m_{K}^{2}-p^{2})(2y_{100}^{r}-4y_{109}^{r}+y_{110}^{r})\right\} \,,
\label{fak}
\ee
where $L^{r}_{i}$ and $y^{r}_{i}$ are the renormalized coupling constants.
Note that the first terms in Eqs. (\ref{fvk}) and (\ref{fak}) correspond to $F_V$ and $F_A$ at $O(p^4)$
\cite{Bijnens93,kl2m},  respectively. We remark that the expressions of Eqs. (\ref{fvk}) and (\ref{fak}) have not been 
explicitly shown in the literature \cite{mod3,mod4}.

\sse{Light Front Quark Model}

In the framework of the LFQM \cite{modLF,lf1,mod1}, the physical accessible kinematics
region is $0\leq p^{2}\leq M_{K}^{2}$ due to the time-like
momentum transfers. The general structure of the phenomenological light front (LF) meson
wave function is based only on the $q\bar{q}$ Fock space sector.
It can be expressed by an anti-quark $\bar{s}$ and a quark $u$
with the total momentum $(p+q)$ such as:
\begin{eqnarray}
|K(p+q)\,\rangle&=& \sum_{\lambda _{1}\lambda_{2}}\int [dk_{1}][dk_{2}]
2(2\pi)^{3}\delta ^{3}(p+q-k_{1}-k_{2})  \nonumber \\
&& ~~~~~~~~ \times \Phi _{K}^{\lambda _{1}\lambda _{2}}(z,k_{\bot})
b_{\bar{s}}^{+}(k_{1},\lambda _{1}) d_{u}^{+}( k_{2},\lambda _{2})
|0\,\rangle\,,
\label{LFw}
\end{eqnarray}
where $\Phi _{K}^{\lambda _{1}\lambda _{2}}$ is the amplitude of the corresponding
$\bar{s}(u)$ and
$k_{1(2)}$ is the on-mass shell LF momentum of the internal
quark. The LF relative momentum variables $(z,k_{\bot})$ are defined by
\begin{eqnarray}
&& k^+_1=z_1 (p+q)^+, \quad k^+_2=z_2 (p+q)^+, \quad z_1+z_2=1, \nonumber \\
&& k_{1\bot}=z_1 (p+q)_\bot+k_\bot, \quad k_{2\bot}=z_2
                (p+q)_\bot-k_\bot\,,
\end{eqnarray}
and
\be
\Phi _{K}^{\lambda _{1}\lambda _{2}}(z,k_{\bot })=\left( \frac{%
k_{1}^{+}k_{2}^{+}}{2[M_{0}^{2}-\left( m_{s}-m_{u} \right) ^{2}]}\right)^{%
\frac{1}{2}}\overline{u}\left( k_{1}, \lambda _{1}\right) \gamma^{5}v\left(
k_{2},\lambda _{2}\right) \phi(z,k_{\bot}) \,,  \label{n6}
\ee
with $\phi(z,k_{\bot})$ being the space part of the wave function, which depends on
the dynamics. The amplitude
of $\phi(z,k_{\bot})$ can be solved in principles by
the LF QCD bound state equation \cite{lf3,lf4}.
However, we use the Gaussian type wave function in this study:
\be
\phi(z,k_{\bot})=N\sqrt{\frac{dk_{z}}{dz}}
\exp \left( -\frac{\vec{k}^{2}} {2\omega_{K}^{2}}\right) \,.
\label{7}
\ee

  From Eqs. (\ref{LFw})-(\ref{7}), the hadronic matrix elements in
Eq. (\ref{4}) are found to be
\be
&&\langle\gamma (q)|\bar{s}\gamma^{\mu }\,(1-\gamma
_{5})\,u|\,K(p+q)\, \rangle=\int \frac{d^{4}k_1'}{(2 \pi)^{4}}
\Lambda_{K} \nonumber \\ &&\times\bigg\{
\gamma_{5}\frac{i(-\strich{k}^{'}_{2}+m_{u})}
{k_{2}^{'2}-m_{u}^{2}+i\epsilon} ie_{u}\strich{\epsilon}
\frac{i(\strich{k}_{1}+m_{u})} {k_{1}^{2}-m_{u}^{2}+i\epsilon}
\gamma^{\mu }(1-\gamma
_{5})\frac{i(\strich{k}'_{1}+m_{s})}
{k_{1}^{'2}-m_{s}^{2}+i\epsilon}
\nonumber \\ &&+(u\leftrightarrow s \, , k'_{1}\,(k_1) \leftrightarrow k'_{2}\,(k_2)
)\bigg\}\,,
\label{int}
\ee
where $\Lambda_{K}$ is a vertex function
related to the $u\bar{s}$ bound state of the $K$ meson, $k_2=q-k_1$
and $k'_1=(p+q)-k'_2=k_1+p$. After integrating over the LF momentum
$k_{1}^{-}$ in Eq. (\ref{int}), we get
\be
&&\langle\gamma (q)|\bar{s}\gamma^{\mu }\,(1-\gamma
_{5})\,u|\,K(p+q)\, \rangle\nonumber \\
&&
=\int_{q}^{p+q} [d^{3}k'_{1}]\bigg\{\frac{1}{k_{1}^{-}-k_{1on}^{-}}
(I^{\mu\nu}|_{k_{1on}^{'-}}) \frac{\Lambda_{P}}{k_{2}^{'-}-k_{2on}^{'-}}
+(u\leftrightarrow s \, , k'_{1}\,(k_1) \leftrightarrow k'_{2}\,(k_2))\bigg\}
\,\,,
\label{22}
\ee
where
\be
&&[d^{3}k'_{1}]=\frac{dk_{1}^{+}dk_{1\bot}}{2(2\pi)^{3} k_1^{'+}
k_2^{'+} k_1^{+}} ~, \nonumber \\
&&I^{\mu\nu}|_{k_{1on}^{-}}=Tr\bigg\{
\gamma_{5}(-\strich{k}^{'}_{2}+m_{u})ie_{u}\strich{\epsilon}
(\strich{k}_{1}+m_{u})\gamma^{\mu } (1-\gamma
_{5})(\strich{k}^{'}_{1}+m_{s})\bigg\}~, \nonumber \\
&&k_{ion}^{-}=\frac{m_{i}^{2}+k_{i\bot}^{2}}{k_{i}^{+}}~,~
k_{1(2)}^{'-}=p_{on}^{-}-k_{2(1)on}^{'-} ~,~
k_{1}^{-}=q^{-}-k_{2on}^{-} \,,
\label{trace}%
\ee
with $\{on\}$ representing the on-shell particles.
For the kaon, the vertex function $\Lambda_{P}$ in Eqs.  (\ref{int}) and (\ref{22}) is
given by
\cite{vex1,vex2}:
\be
\frac{\Lambda_{P}}{k_{2}^{'-}-k_{2on}^{'-}}\to
{\sqrt{k_1^{'+} k_2^{'+}}\over \sqrt{2} ~{\widetilde M_0}}~ \phi(z',
k_\bot)\,.
\ee %
To calculate the right hand part of Eq. (\ref{22}), we choose
a frame with the transverse momentum $p_{\bot}$ = $0$ so that
$p^{2}=p^{+}p^{-} \geq 0$ covers the entire range of the momentum
transfers. Here, we have used the LF momentum variables $(x,k_{\bot})$.  Hence, the relevant
quark momentum variables in Fig. 1 are
\be
&k_{1}^{'+}=(1-z')(p+q)^+,
~k_{2}^{'+}=z'(p+q)^+,
~k_{1\perp}^{'}=(1-z')q_\perp+k_\perp^{'},~k_{2\perp}^{'}=z'q_\perp-k^{'}_\perp\,,
\nonumber \\
&k_{1}^{+}=(1-z)q^+,~~k_{2}^{+}=zq^+,
~~k_{1\perp}=(1-z)q_\perp+k_\perp,~~k_{2\perp}=zq_\perp-k_\perp\,.
\label{transmom}
\ee
By considering the good component as ``$\mu=+$'', the hadronic
matrix elements in Eq. (\ref{4}) can be rewritten as:
\be
\langle\gamma (q)| s_{+}^{+}\gamma_{5}u_{+}|K(p+q)\,\rangle
        &=&-e\frac{F_{A}}{2m_{K}}\left( \epsilon
        _{\bot }^{*}\cdot q_{\bot }\right) p^{+}\,,
                \nonumber \\
\langle\gamma (q)|s_+^+u_+|K(p+q)\,\rangle&=&-ie\frac{F_{V}}{2m_{K}}
        \epsilon ^{ij}\epsilon _{i}^{*}q_{j}p^{+}\,.
                \label{ff}
\ee
Using  Eq,~(\ref{transmom}), the trace part $I^{\mu\nu}$ in
Eq.~(\ref{trace}) can be carried out.
By comparing Eq. (\ref{22}) with Eq. (\ref{ff}), we obtain
 the form factors $F_{V,A}$
  to be:
\be
F_{A}(p^2) &=&4m_{K}
        \int \frac{dz\,d^{2}k_{\bot }}{2(2\pi)^{3}}\Phi
        \left( z',k_{\bot }^{2}\right) {1\over 1-z'}
                \nonumber \\
&&~~~~~~~~ \times \left\{\frac{2}{3}\frac{m_{u}-Ak_{\bot }^{2}\Theta }
        {m_{u}^{2}+k_{\bot }^{2}}+ \frac{1}{3}\frac{m_{s}+Bk_{\bot }^{2}
        \Theta}{m_{s}^{2}+
        k_{\bot}^{2}}  \right\}\,,   \nonumber \\
F_{V}(p^2) &=&8m_{K}
        \int \frac{dz\,d^{2}k_{\bot }}{2\left( 2\pi \right) ^{3}}\Phi
        \left( z',k_{\bot }^{2}\right) {1\over 1-z'}
                \nonumber \\
&&\left\{ \frac{2}{3}\frac{m_{u}-
        z'\left( m_{s}-m_{u}\right) k_{\bot }^{2}\Theta }{m_{u}^{2}
        +k_{\bot }^{2}}-\frac{1}{3}\frac{m_{s}+(1-z')(m_{s}-m_{u}) k_{\bot }^{2}
        \Theta }{m_{s}^{2}+k_{\bot }^{2}} \right\}\,,
\label{FFLFQM}
\ee
where
\be
A &=& (1-2z) z'(m_s-m_u) -2 z m_u\,, \nn\\
B &=& (1-2z) z'm_s+m_s+(1-2z) (1-z')m_u\,,  \nn\\
\Phi (z,k_{\bot}^2) &=& N\left( {\frac{z(1-z) }{2[M_0^2-(m_s-m_u)^2]}}%
\right)^{1/2} \sqrt{{\frac{dk_{z}}{dz}}}\exp \left( -{\frac{\vec{k}^{2}}{%
2\omega_K^2}}\right)\,,  \nonumber \\
\Theta &=& {\frac{1}{\Phi(z,k_{\bot}^2) }} {\frac{d\Phi(z,k_{\bot}^{2})}{%
dk_{\bot}^2}} \, ,  \nonumber \\
z^{\prime}&=&z\left(1-{\frac{p^2}{M_K^2}}\right),\
\vec{k}=(\vec{k}_{\bot}, \vec{k}_{z}) \,,
\nn\\
N&=&4 \left({\frac{\pi%
}{\omega_{K}^{2}}}\right)^{\frac{3}{4}}\, ,  ~~
k_{z}= \left( z-\frac{1}{2}\right) M_{0}+\frac{m_{s}^{2}-m_{u}^{2}}{%
2M_{0}} \, , \nonumber \\
M_0^2&=&{\frac{k^2_{\bot}+m_u^2}{z}}+{\frac{k^2_{\bot}+m_s^2}{%
1-z}} \, .
\ee

\se{Differential Decay Rate}

In the $K^{+}$ rest frame, the partial decay rate for $K^{+}\to e^+\nu_e\gamma$ is given by
 \cite{pdg}
\be
d\Gamma =\frac{1}{(2\pi )^{3}}\frac{1}{8m_{K}}\mid M\mid ^{2}dE_{\gamma
}dE_{e}\,,
\label{42}
\ee
where $E_\gamma$ and $E_e$ are photon and electron energies, respectively.
To describe the kinematics of $K^{+}\to e^{+}\nu_{e}\gamma$, we introduce two dimensionless
variables,
defined by $x=2E_{\gamma}/m_{K}$ and $y=2E_e/m_{K}$,
with their  physical allowed regions being
\be
0 &\leq &x\leq 1-r_e \,,
\nonumber \\
1-x+\frac{r_e}{1-x} &\leq &y \leq 1+r_e\,,
\ee
where $r_e=m_{e}^{2}/m_{K}^{2}$.
The relation between the transfer momentum $p^2$ and $x$ is given by:
\be
p^2=m_{K}^{2}(1-x).
\ee
  From Eqs . (\ref{eqn:sd})
and  (\ref{42}), we obtain the double differential decay rate of $K^{+}\to e^{+}\nu_{e}\gamma$ as
\be
\frac{d^{2}\Gamma }{dx\, dy }
&=&
\frac{m_{K}^{5}}{64\pi
^{2}} \alpha G_{F}^{2}\sin^{2}\theta_c (1-\lambda ) A(x,y),
\label{Rate}
\ee
where  $\lambda=(x+y-1-r_e)/x$ and
\be
A(x,y)&=&A_{IB}(x,y)+A_{SD^+}(x,y)+A_{SD^-}(x,y)+A_{INT^+}(x,y)+A_{INT^-}(x,y)\,,
\nn
\label{43}
\ee
\be
A_{IB}(x,y) &=&
{\frac{4r_{e}|F_{K}|^{2}}{m_K^2\lambda x^{2}}}
\left[ x^{2}+2(1-r_e)
\left(1-x-{\frac{r_e}{\lambda}}\right)\right]\,,
  \nonumber \\
A_{SD^+}(x,y) &=&
 |F_{V}+F_{A}|^{2}
{\frac{x^{2}\lambda ^{2}}{1-\lambda }}
\left( 1-x-{\frac{r_{e}}{\lambda }}\right) \,,  \nonumber \\
  A_{SD^-}(x,y) &=&
 |F_{V}-F_{A}|^{2}x^{2}(y-\lambda)\,,
  \nonumber \\
A_{INT^+}(x,y) &=&
-{4r_{e}\over m_K} Re[F_{K}(F_{V}+F_{A})^{*}]\left( 1-x
-{\frac{
r_e}{\lambda }}\right) \,,   \nonumber \\
A_{INT^-}(x,y) &=&
{4r_{e}\over m_K}Re[F_{K}(F_{V}-F_{A})^{*}]{\frac{1-y+\lambda }{\lambda }} \,.
\label{48}
\ee
By integrating out the $y$ variable in Eq. (\ref{Rate}), we obtain the differential decay rate as a function of $x$ to be
\be
\frac{d\Gamma }{dx}
&=&
\frac{m_{K}^{5}}{64\pi
^{2}} \alpha G_{F}^{2}\sin^{2}\theta_c  A(x)
\label{Rate1}
\ee
where
\be
A(x)&=&A_{IB}(x)+A_{SD^+}(x)+A_{SD^-}(x)+A_{INT^+}(x)+A_{INT^-}(x)\,,
\nn
\ee
\be
A_{IB}(x)
 &=& {4r_eF_K^2\over m_K^2}
\left[{(x+r_e-1)[x^2+4(1-r_e)(1-x)]\over 1-x} \right.
\nn\\
&&~~~~~~~~~~\left.-
{\frac{x^{2}+2(1-r_{e})(1-x+r_{e})}{x}}\ln{\frac{r_{e}}{1-x}}
\right] \,, \label{n49}
\nn\\
A_{SD^+}(x) &=&
|F_{V}+F_{A}|^{2}x^{3}\left[ { 1-x\over 3}-{r_e\over 2}
  +{r_e^3\over 6(1-x)^2}\right]\,,
 \nn\\
 A_{SD^-}(x) &=&
 |F_{V}-F_{A}|^{2}x^{3}\left[ { 1-x\over 3}-{r_e\over 2}
  +{r_e^3\over 6(1-x)^2}\right]\,,
\nn \\
A_{INT^+}(x)&=& {4r_e\over m_K}Re[F_{K}(F_{V}+F_{A})^{*}] x \left[
\frac{1-x}{2}-{\frac{ r_{e}^{2}}{2(1-x)}} +
r_{e}\ln{\frac{r_e}{1-x}}\right]\,,
\nn\\
A_{INT^-}(x)&=& {4r_e\over m_K}Re[F_{K}(F_{V}-F_{A})^{*}] x \left[
{\frac{-1+3x}{2}}+{\frac{ r_e^2-2 xr_e}{2(1-x)}}
+\left(x-r_{e}\right)\ln{\frac{r_e}{1-x}} \right]\,. \label{n51}
\ee It is clear that the contributions to the decay rate from the
$IB$ and $INT^\pm$ parts are suppressed due to the small electron
mass.

\begin{figure}[htbp]
\includegraphics*[width=4in]{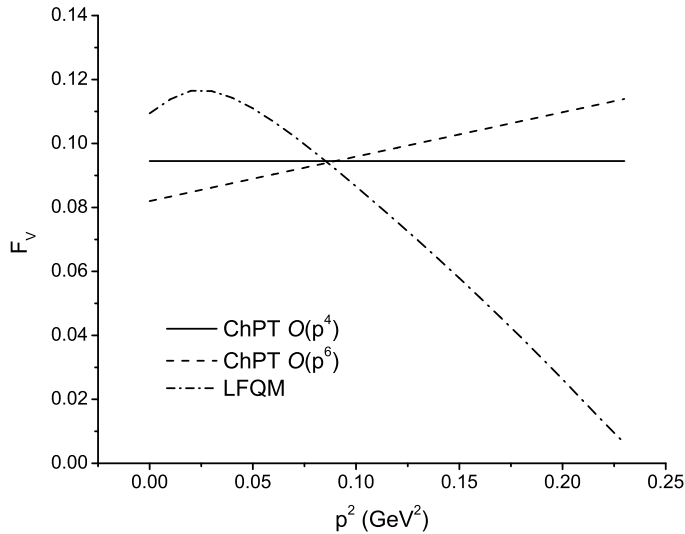}
\caption{ $F_V(p^2)$ as a
function of the transfer momentum $p^2$. }
\label{F2}
\end{figure}

\begin{figure}[htbp]
\includegraphics*[width=4in]{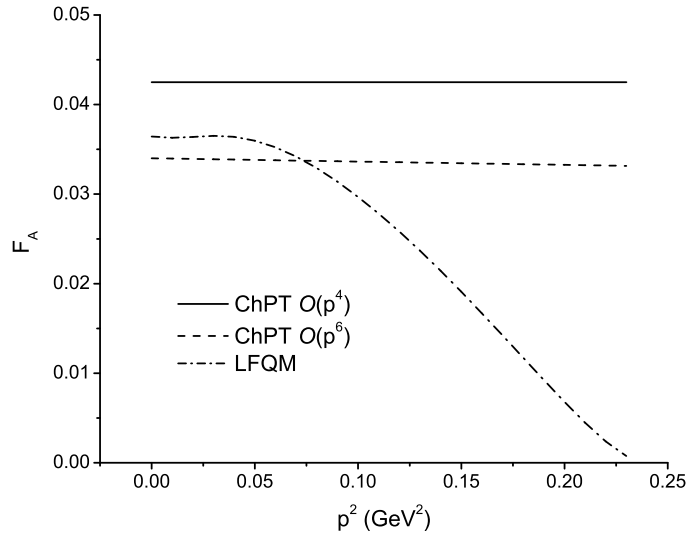}
\caption{ $F_A(p^2)$ as a
function of the transfer momentum $p^2$. }
\label{F3}
\end{figure}

\se{Numerical results}
The numerical values of $F_{A,V}(p^2)$ in the ChPT of $O(p^6)$
have been shown in Figs. 5 and 6 of Ref. \cite{mod4}.
To compare these values with those in the LFQM, we
plot the results in Figs. \ref{F2} and \ref{F3}. In these figures, we have also included the results in the ChPT at $O(p^4)$.
For the calculations of the ChPT \cite{mod4}, we have taken
$m_K=0.495$, $m_\pi=0.14$, $m_\eta=0.55$ and  $m_\rho=0.77$, $F_K=0.112$ GeV
and the  renormalized coefficients of
$(L^r_1,L^r_2,L^r_3,L^r_9,L^r_{10})$, $(C^r_7,C^r_{11},C^r_{22})$ and
$(y^r_{100},y^r_{104},y^r_{109},y^r_{110})$ to be $(0.53,0.71,-2.72,6.9,-5.5)\times 10^{-3}$ \cite{33},
$(0.013,-6.37,6.52)\times 10^{-3}\,GeV^{-2}$ \cite{29} and $(1.09,-0.36,0.40,-0.52)\times 10^{-4}/F_K^2$ \cite{25}, respectively. For some other possible sets of coefficients, see Ref. \cite{mod4} as well as the recent review in
Ref. \cite{Bi2007}. We note that we have ignored the contributions from $p^2$-nondependent terms involving $y^r_i$.
On the other hand,
 the $p^2$-dependence of $F_A(p^2)$ for the ChPT at
$O(p^6)$  are insensitive due to the small contributions related to $y^r_i$ \cite{mod4}.
We emphasize that as illustrated in Figs. \ref{F2} and \ref{F3},
   the form factors $F_{V,A}$ at $O(p^4)$ in the ChPT are constants \cite{Bijnens93}.
To evaluate the form factors of $F_{V,A}$ from Eq. (\ref{FFLFQM})
in the LFQM , we have used $m_{u}=0.26$,
$m_{s}=0.37$ and $\omega_K =0.382$ in $GeV$.
In Table \ref{Table1}, we explicitly display the values of $F_{V,A}(p^2=0)$.
\begin{table}[h]
    \caption{
 The form factors of $F_{V}(0)$ and $F_{A}(0)$
in (a) the ChPT at $O(p^4)$ \cite{Bijnens93}, (b) the ChPT of $O(p^6)$ and (c) the LFQM.}
 \vskip 0.2in
 \label{Table1}
\begin{tabular}{|c||c|c|c|} \hline
Model & $F_{V}(0)$ & $F_{A}(0)$
\\ \hline \hline
(a) & 0.0945 & 0.0425
\\ \hline
(b) & 0.082 & 0.034
\\ \hline
 (c) & 0.106 & 0.036
\\ \hline
\end{tabular}
 \end{table}

By integrating out the variable $x$  in Eq. (\ref{Rate1}), in Table \ref{Table2} we give the
decay branching ratio of $K^+ \to e^+ \nu_e \gamma$
in (a) the ChPT at $O(p^4)$, (b) the ChPT of $O(p^6)$ and (c) the LFQM.
Here, as
the IB term diverges at the limit of
$x\to 0$ corresponding to $p^2\to p^2_{max}=m_K^2$, we have used the cuts of $x=0.01$ and $0.1$, respectively.
With the cuts, from Table \ref{Table2} we see that
both IB and INT$^\pm$ contributions are much smaller than
 the SD$^\pm$ ones, which are insensitive to the cut.
 We remarks that in Table \ref{Table2}, our results for the SD$^+$ contribution to the decay branching ratio
 in the ChPT of $O(p^6)$ and LFQM are $1.15$ and $1.12\times 10^{-5}$, which are smaller than 
that of $1.52\pm0.23\times 10^{-5}$  \cite{75,79}
quoted by the PDG \cite{pdg},
 respectively. Note that the value in the  PDG was based on the combination of the data in Refs. \cite{75} and 
 \cite{79}, in which large constant values of $F_A+F_V=0.150^{\small +0.018}_{\small -0.023}$ and
 $0.147\pm 0.011$ were used, respectively. It is clear that to compare the data with the theoretical predictions,
 proper form factors should be used in the data analysis.
 
\begin{figure}[htbp]
\includegraphics*[width=4in]{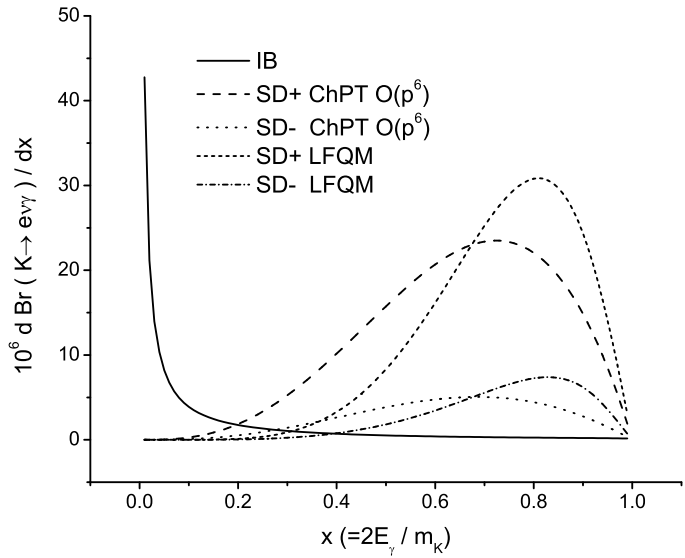}
\caption{The IB and SD$^\pm$
parts of the differential decay branching ratio
as  functions of $x=2E_{\gamma}/m_{K}$.}
\label{F4}
\end{figure}

\begin{figure}[htbp]
\includegraphics*[width=4in]{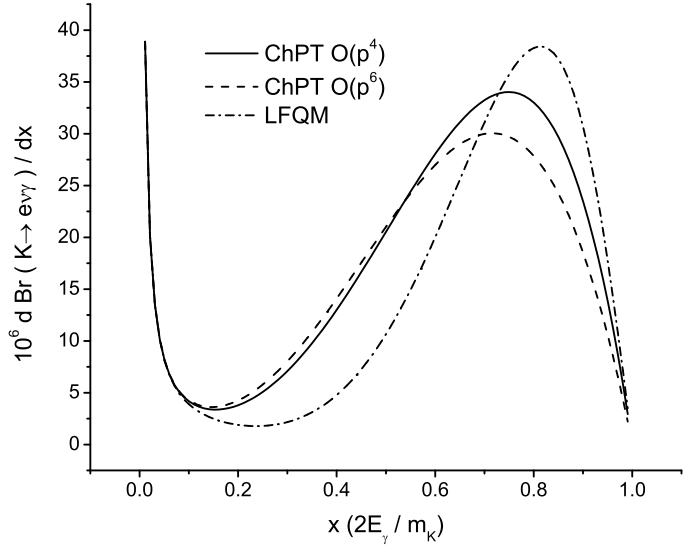}
\caption{The
differential decay branching ratio
as a function of $x=2E_{\gamma}/m_{K}$.}
\label{F5}
\end{figure}

To show the behaves of the various contributions in the ChPT and LFQM,
we present  the IB and
SD$^\pm$
parts of the differential decay
branching ratio as functions of  $x$ in Fig.~\ref{F4}.
 Here, we do not plot the INT$^\pm$ contributions in Fig.~4 as they are vanishingly small.
As shown in the figure, in the
small $x$ region
 there is an enhancement for the IB part,
whereas those from the
SD$^\pm$
parts are close to zero.
 In Fig. \ref{F5}, we also display the spectrum of the differential decay
branching ratio vs. $x$ in the ChPT at both $O(p^4)$ and $O(p^6)$ and the LFQM.

\begin{table}[htbp]
   \caption{
 The decay branching ratio  of $K^+ \to e^+ \nu_e \gamma$ (in units of $10^{-5}$)
in (a) the ChPT at $O(p^4)$, (b) the ChPT of $O(p^6)$ and (c) the LFQM with the cuts of $x=0.01$ and $x=0.1$, respectively.}
 \vskip 0.2in
\label{Table2}
\begin{tabular}{|c|c||c|c|c|c|c|c|c|} \hline
Model &Cut & IB & SD$^+$ & SD$^-$ &  INT$^+$ &  INT$^-$ & Total
\\ \hline \hline
(a) & $x=0.01$&$ 1.65\times 10^{-1} $
& $ 1.34 $ & $1.93\times 10^{-1} $ & $6.43\times 10^{-5} $
& $-1.10\times 10^{-3}$  & $ 1.70 $
\\
 &$x=0.1$& $ 0.69\times 10^{-1} $
& $ 1.34 $ & $1.93\times 10^{-1} $ & $6.43\times 10^{-5} $
& $-1.10\times 10^{-3}$  & $ 1.60 $
\\ \hline
(b) &$x=0.01$& $ 1.65\times 10^{-1} $
& $ 1.15 $ & $2.58\times 10^{-1} $ & $6.22\times 10^{-5} $
& $-1.21\times 10^{-3}$  & $ 1.57 $
\\
& $x=0.1$&$ 0.69\times 10^{-1} $
& $ 1.15 $ & $2.58\times 10^{-1} $ & $6.22\times 10^{-5} $
& $-1.21\times 10^{-3}$  & $ 1.47 $
\\ \hline
 (c) & $x=0.01$&$ 1.65\times 10^{-1} $
& $ 1.12 $ & $2.59\times 10^{-1} $ & $4.33\times 10^{-5} $
& $-1.29\times 10^{-3}$  & $ 1.54 $
\\
 & $x=0.1$&$ 0.69\times 10^{-1} $
& $ 1.12 $ & $2.59\times 10^{-1} $ & $4.33\times 10^{-5} $
& $-1.29\times 10^{-3}$  & $ 1.44 $
\\ \hline
\end{tabular}
\end{table}

 From Fig. \ref{F5},
we see that in the region of $x<0.7$ or $E_\gamma<173$ MeV,
 the decay branching ratio in the LFQM is much smaller
than that in the ChPT at $O(p^6)$. On the other hand, in the region of $x>0.7$
the statement is reversed. However, if we only consider the contributions in the ChPT at $O(p^4)$,
the conclusion is weaker. In Table \ref{Table3}, we illustrate the decay branching ratio in the regions of
$0.1<x<0.7$ and $0.7<x<1$ from the various approaches, respectively.
\begin{table}[htbp]
   \caption{
 The decay branching ratio  of $K^+ \to e^+ \nu_e \gamma$ (in units of $10^{-5}$)
in  the regions of
$0.1<x<0.7$ and $0.7<x<1$ from the various approaches, respectively.}
 \vskip 0.2in
\label{Table3}
\begin{tabular}{|c||c|c|c|} \hline
Region & ChPT of $O(p^4)$ & ChPT of $O(p^6)$ & LFQM
\\ \hline \hline
 $0.1<x<0.7$ & $0.871$ & $0.871$ & $0.541$
 \\ \hline
 $0.7<x<1$ & $0.733$ & $0.606$ & $0.902$
\\ \hline
\end{tabular}
\end{table}
The main reasons for the differences are due to the form factors.
The form factors of the ChPT at  $O(p^4)$ are constant
and straight lines at  $O(p^6)$, whereas  in  the LFQM they
 are the overlap between the wave functions of the $K$
meson and photon and become zero  when $x\to 0$ or $p^2\to p^2_{max}=m_K^2$.
It is clear in the future data analysis such as the one at the experiment BNL-E949 \cite{E949},
one could concentrate on these two regions to find out which model is preferred.

\section{Conclusions}

We have studied the axial-vector and vector form factors  of the $K ^+\to \gamma$ transition
in the LFQM and   ChPT of $O(p^6)$.
Based on these form factors, we have calculated the decay branching ratio of
 $K^+ \to e^+ \nu_{e}\gamma$. We have demonstrated that the SD part gives the dominant contribution
  to the decay in the whole allowed region of the photon energy except the low endpoint.
  Explicitly, we have found that, in the SM with the cut of $x=0.01$ ($0.1$),
 the decay branching ratio of $K^+ \to e^+ \nu_{e}\gamma$ is
 $1.54\ (1.44)\times 10^{-5}$ and $1.57\ (1.47)\times 10^{-5}$
 in the LFQM and ChPT,
respectively.   Future precision experimental measurements on the decay spectrum \cite{E949}
should give us some useful information to
  determine the SD contribution as well as the vector and axial-vector form factors.

\section*{Acknowledgments}
We would like to thank Professor D. A. Bryman
 for the suggestion of studying the decay spectrum  and many
  useful
discussions.
This work is supported in part by the National Science Council of
R.O.C. under Contract
 \#s: NSC-95-2112-M-006-013-MY2,
  NSC-95-2112-M-007-059-MY3 and
  NSC-95-2112-M-277-001.

\end{document}